\documentclass[english,a4paper]{IEEEconf}
\usepackage[latin9]{inputenc}
\usepackage{prettyref}
\usepackage{graphicx}

\usepackage{listings}

\lstdefinelanguage{utc}{morekeywords={thread,index,create,sync,shared,place,family,DEFAULT,LOCAL,get_ncores}}
\usepackage{babel}

\begin{document}

\title{Extending and Implementing the Self-adaptive Virtual Processor for Distributed Memory Architectures}

\author{Michiel W. van Tol\\
\begin{affiliation}
University of Amsterdam\\
Amsterdam, The Netherlands
\end{affiliation} \\
\email{mwvantol@uva.nl}
\and 
Juha Koivisto\\
\begin{affiliation}
VTT Technical Research Centre of Finland\\
Espoo, Finland
\end{affiliation} \\
\email{Juha.Koivisto@vtt.fi}
}

\maketitle

\begin{abstract}
Many-core architectures of the future are likely to have distributed
memory organizations and need fine grained concurrency
management to be used effectively. The Self-adaptive Virtual Processor
(SVP) is an abstract concurrent programming model which can provide
this, but the model and its current implementations assume a single 
address space shared memory. We investigate and extend SVP to handle 
distributed environments, and discuss a prototype SVP implementation which 
transparently supports execution on heterogeneous distributed memory clusters
over TCP/IP connections, while retaining the original SVP programming model. 
\end{abstract}

\section{Introduction}

As processor architectures are moving into the many-core era, potentially
scaling up to more than 1000s of cores on a chip \cite{Manycore07,IntelTerascale07},
it becomes infeasible to maintain a memory model which guarantees
system-wide sequential consistency. Full cache-coherence will not
scale for such architectures \cite{Heiser2009,Kurian2010} or will
suffer from large latencies, so future many-core architectures are
likely to have a more distributed and weakly consistent memory design.
For example, these could be organized in a similar way as the experimental
48-core Intel SCC research chip \cite{IntelSCC10}, on which each
processor can access both a private and shared memory, but no hardware
cache coherence is provided. In order to exploit many-cores to their
full potential, it is essential to be able to create parallelism at
a fine granularity in order to expose the maximum amount of concurrency.
We require a programming model to express this concurrency, but which
can also handle such distributed memory organizations efficiently.

In this report, we apply and adapt the definition of the Self-adaptive Virtual 
Processor (SVP) to distributed memory organizations, naming this extension 
DSVP\footnote{We use the name DSVP  
throughout the report for matters specific to our extension, and SVP 
for anything that applies to both the original and extended model.}. 
SVP is an abstract concurrent programming and machine model \cite{SVP08},
which evolved from the earlier work on the Microthread CMP architecture
\cite{Microthread06}. It can be used to express concurrency at many
levels of granularity for multi- or manycore systems, and uses weakly
consistent shared memory semantics. As SVP is a generic model to program parallel
systems, this method can be applied to the whole spectrum of memory
organizations; from cc-NUMA machines where you want to maintain locality,
to non cache coherent shared memory machines such as the Intel SCC
or other future many-core architectures, and even a cluster of
nodes on a network, i.e. a heterogeneous distributed system. This
is achieved by extending SVP implementations and the way they are programmed 
to support 
distributed memory spaces, and by translating SVP actions into messages
in a distributed environment. Using this approach, we believe that
we have made a step forward in efficiently targeting any architecture 
within the aforementioned spectrum.

In order to go into further details of this work, we will give a short
introduction to the semantics and actions of the SVP model (\prettyref{sec:Background}),
and we describe its current memory consistency model. We then define how
we can apply SVP to a distributed environment, at which level of granularity 
we can identify and distribute software components, and how we identify 
their dependencies in order to communicate data between nodes in 
\prettyref{sec:Model-and-Language}.
We then discuss our research prototype that implements these
techniques using messages over TCP/IP in \prettyref{sec:Implementation}, 
and show that this follows the original SVP memory consistency model.
This implementation is then evaluated and discussed (\prettyref{sec:Evaluation}),
where we show that this approach integrates nicely with SVP as SVP
actions are handled transparently and uniformly between local and
remote executions. This discussion is continued in \prettyref{sec:Related-Work}
where we compare it with a broad spectrum of related approaches in
distributed computing. We then conclude in \prettyref{sec:Conclusion}.

\section{The SVP Model}\label{sec:Background}

SVP is a generic concurrent programming and machine model \cite{SVP08},
of which both coarse \cite{SVP-PTL09} and fine \cite{Microgrid09}
grained implementations are available. The goal of SVP is to be able
to express concurrency, without having to explicitly manage it. The
$\mu$TC language \cite{uTC06}, based on C99, has been defined to
capture the semantics of SVP. This language is used to drive several
SVP implementations, as it extends traditional C with syntax to express
all SVP actions.

The SVP model defines a set of actions to express concurrency on groups
(families) of indexed identical threads. Each thread can execute a
\emph{create} action to start a new concurrent child family of threads,
and later on use the \emph{sync} action to wait for its termination,
implementing a fork-join style of parallelism. The \emph{create} action
has a set of parameters to control the number and sequence of created
threads, as well as a reference to the thread function that the threads
will execute. This thread function can have arguments, defined by
SVP's communication channels explained later on.

As any thread can create a new family, the concurrency in a program
can consist of many hierarchical levels, often referred to as the
\emph{concurrency tree} of a program. Besides these two basic constructs,
there is the \emph{kill} action to asynchronously terminate an execution.

\paragraph{Resources}

SVP code has no notion of what resources are, as it is resource and scheduling
naive. However, the concept of \emph{place} is provided as an abstract
resource identifier. On a \emph{create} action a \emph{place} can
be specified where the new family should be created, binding the execution
onto a certain resource. What this \emph{place} physically maps to,
is left up to the SVP implementation; for example, on a many-core
architecture like the Microgrid, it could be a group of processors.
On other implementations it could, for example, be a reserved piece
of FPGA fabric, an ASIC, or some time-sliced execution slot on a single-
or multi- processor system. As long as the underlying implementation
supports it, multiple \emph{places} can be virtualized onto a single
resource.

There is one important property that a \emph{place} can have; it can
be \emph{exclusive}. This means that each \emph{create} on such an
\emph{exclusive} \emph{place} will be sequentialized. Only one family
can be executing on such a place at a time, providing us with a mutual
exclusion mechanism.

\paragraph{Communication and Synchronization}

Synchronized communication is provided through a set of channels,
which run between threads in a family and their parent thread. There
are two types of unidirectional write-once channels; \emph{global}
and \emph{shared} of which multiple can be present. These channels
have non-blocking writes and blocking reads. A \emph{global} channel
allows vertical communication in the concurrency tree from the parent
thread to all threads in the family. A \emph{shared} channel allows
horizontal communication, as it daisy-chains through the sequence
of threads in the family, connecting the parent to the first thread
and the last thread back to the parent. These channels are defined
as arguments of a thread function, similar to normal function arguments,
and identify the data dependencies between the threads. Due to this
restricted definition, and under restricted use of exclusive places,
we can guarantee that the model is composable
and free of communication deadlock~\cite{FormalSVP07}. Furthermore,
this implies that every family of threads has a very well defined
sequential schedule if concurrent execution is infeasible, as it is
guaranteed that a family can run to completion when all of its threads
are executed in sequence. This enables program transformations that
sequentialize families into loops at the leaves of the concurrency
tree, allowing us to adapt the granularity and amount of exposed concurrency
in an SVP program for a specific platform.

\paragraph{Memory Consistency}

The model assumes a global, single address space, shared memory. However,
this is seen as asynchronous and has a restricted consistency model.
Therefore it is not suitable for synchronizations, and no explicit
memory barriers or atomic operations are provided. The consistency
model is described by the following three rules:
\begin{itemize}
\item Upon creation, a child family is guaranteed to see the same memory
state as the parent thread saw at the point where it executed \emph{create}.
\item The parent thread is guaranteed to see the changes to memory by a
child family only when \emph{sync} on that family has completed.
\item Subsequent families created on an \emph{exclusive place} are guaranteed
to see the changes to memory made earlier by other families on that
place.
\end{itemize}
The memory consistency relationship between parent and child threads
somewhat resembles the well-known \emph{release consistency} model
\cite{ReleaseConsistency90}. In that sense, the point of \emph{create
}resembles an \emph{acquire}, and the point of \emph{sync} resembles
the \emph{release}. We should note that the third rule is a
very important property as it can be used to implement communication
between two arbitrary threads, but it can also be used to implement
a service; state is resident at the \emph{exclusive place} and instances
of the functions implementing that service are created on the \emph{place}
by its clients. An example of such a service has been presented in
\cite{SVPIO10} for the SVP based Microgrid architecture.

Data passed through the \emph{global} or \emph{shared} channels
is always considered consistent. However, it is likely that in certain
implementations the channels are limited to only scalar values, therefore
a reference to a datastructure in memory would be passed instead of the
structure itself. An implementation then has to guarantee that there
is memory consistency for the referenced structure when it is read from
the channel.

\paragraph{Example}

\begin{figure}[ht]
\begin{center}
\includegraphics[width=0.85\columnwidth]{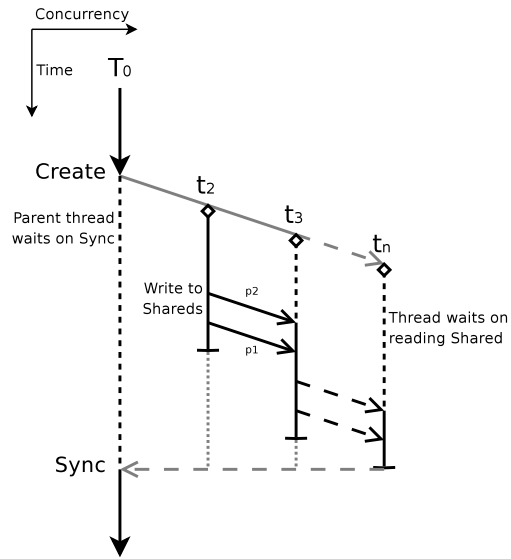}
\end{center}
\caption{\label{fig:Fibonacci-diagram}Fibonacci time-concurrency diagram}
\end{figure}

The basic concepts of SVP are illustrated in Figure \ref{fig:Fibonacci-code} 
and Figure \ref{fig:Fibonacci-diagram}
using some example code that generates a Fibonacci sequence and stores
it in an array. It must be noted that this example yields little exploitable
concurrency, but is merely used as a simple illustration of the concepts.

\begin{figure}[ht]
\begin{lstlisting}
thread fibonacci(shared int p1, 
        shared int p2, int* result)
{
    index i;
    result[i] = p1 + p2;
    p2 = p1;
    p1 = result[i];
}

main()
{
    family  fid;
    place   pid = PLACE_DEFAULT;
    int     result[N];
    int a = result[1] = 1;
    int b = result[0] = 0;
    
    create(fid;pid;2;N;;) 
        fibonacci(a, b, result);
    sync(fid);
}

\end{lstlisting}
\caption{\label{fig:Fibonacci-code}Fibonacci code example}
\end{figure}

In Figure \ref{fig:Fibonacci-code} we show the C-like $\mu$TC code that
implements Fibonacci, with the iterations of the algorithm defined as a thread
function in lines 1-8. The definition on lines 1 and 2 identifies the shared 
channels for the two dependencies in Fibonacci, as well as a global that will
pass the pointer for the result array. The shared channels are read
implicitly on line 5, and written on lines 6 and 7. Line 10 to 21 show
the main function of the program that will start the concurrent Fibonacci
iterations. Line 12 defines a variable that can hold a family identifier 
which is set by the \emph{create} on line 18. Line 13 defines 
a place identifier which is set to a default defined by the SVP implementation. 
Then the initial values for the algorithm are set in lines 15 and 16, and the 
spawn of concurrent iterations is done with the \emph{create} statement in 
lines 18 and 19 creating a family of indexed threads from 2 to \emph{N} on the 
place identified by pid. The two omitted parameters can be used to further 
control the creation and indexing of threads by step and block size. 
Information to identify the created family is stored in \emph{fid}, and the 
\emph{sync} statement on line 20 uses this to have the main thread wait until 
all threads in the Fibonacci family have terminated. On line 19, the variables 
\emph{a} and \emph{b} are used to initialize the shared channels for the 
fibonacci family, providing the values that the first thread will read, as 
well as the pointer to the array to store the results.

In Figure \ref{fig:Fibonacci-diagram} the time-concurrency diagram is shown
that corresponds with our example, which shows the interactions between 
threads. $T_{o}$ is the main thread that executes the \emph{create},
which then waits immediately using \emph{sync} on the termination
of the created family of threads. The fibonacci threads $t_{2,}t_{3}...t_{n}$
are then started, and all but the first will immediately block and
suspend on reading the shared channels. The first thread that received the 
shared values from the parent can execute, and then passes on the values to 
the next thread. As Fibonacci requires the value of the $n-1$th and the $n-2$th
iteration, the value from the shared channel \emph{p1} is forwarded to 
\emph{p2} in each thread.
Only when its shareds are written, a suspended thread will continue
its execution again. When all threads have completed, the \emph{sync}
in the parent thread completes and it resumes its execution and can
now safely use the results array. The writes to \emph{p1} and \emph{p2}
by the last thread could be read by the parent again after the \emph{sync},
but are not used in this example.

\section{Distributed SVP}\label{sec:Model-and-Language}

As we have claimed in the introduction, the work
described here can be applied to a whole range of possible target
architectures, we require a definition of what the \emph{distributed
environment} is that we want to apply SVP to, and how we represent
this in the model. Then we will discuss how we identify the software
components that we want to distribute, and how we identify which data
to communicate.

\paragraph{Distributed Environment in SVP}

We define our distributed environment to consist of a set of processing
resources which implement SVP, be it either in software or in hardware,
and that are grouped into nodes of one or more of these resources.
We define a node to have a single addressable, coherent, and optionally
uniform, access to some memory. The nodes are interconnected by an
infrastructure consisting of one or more, possibly heterogeneous,
networks, on which each node can, directly or indirectly, send a message
to any other node. A \emph{place} is identified as a subset of one
or more (or all) resources within a single node, which therefore inherits
the properties that we have just described.

To give some more concrete examples; in a NUMA system which is not fully
cache coherent, a node would be a group of processors that are in
a single NUMA domain that is internally cache coherent. A \emph{place} 
would then be one or more of these processors. In the case of a networked 
(e.g. Ethernet) cluster of multi-core machines, each machine would be a node 
and each core in a machine could be identified by its own \emph{place}. However,
if these multi-core machines would be cache coherent NUMA architectures
themselves, one could optionally choose to subdivide these into separate
nodes per NUMA domain to be able to express and exploit memory locality.
As a final example, the Intel SCC \cite{IntelSCC10} does not provide
any cache coherence, so a node and a \emph{place} would be only a single 
core on the chip.

It should be noted that within a single node, the classic definition
of SVP works perfectly, and we only need to take into account interactions
that are \emph{remote}, i.e. that are between nodes, in order
to apply it to a distributed environment. All SVP actions can be trivially
translated into messages that can be sent across a network, and the
place concept is nicely suited to capture the necessary addressing
information on which node this place is physically located. By using
a place on a remote node, a create transparently turns from a local
concurrency control into a concurrent remote procedure call. Threads
in a family created this way can then again create more families there
locally, or at some point decide to distribute their child families
to other nodes again. However, the challenge lies in defining a way
to handle a distributed memory organization instead of a loosely shared
memory system. We need to define how, and at which level of granularity,
we can identify parts of our program that we can distribute to other
nodes.

\paragraph{Software Components}

Using the restrictions that SVP imposes, we can make some assumptions
about the structure of SVP programs. Because a program is structured
as a hierarchical tree of concurrency, most computation, and therefore
data production and/or consumption, takes place at the more fine grained
concurrency in the outer branches and leaf nodes in the tree. An application
can be seen as a collection of software components at the outer branches,
connected together with control code higher up the hierarchy. Due
to the restrictions in communication and synchronization that SVP
imposes, we can assume that these software components are relatively
independent, and therefore are very suitable for distribution across
different nodes.

Having this view in mind, and by taking the memory consistency model
defined previously, we can make some further assumptions about the
communication of data within an SVP program. As communication is restricted
at the family level, where a thread can communicate data through the
\emph{shared} and \emph{global} channels to a family it creates, we
can make the following observation; The created threads will, disregarding
global references, only access data that is either passed through
these \emph{shared} or \emph{global} channels, or data in memory that
is accessed through a reference that is passed this way. Newly generated
data that needs to be communicated back to the parent, has to be passed
back again through the \emph{shared} channel. Therefore, the data
dependencies of software components are identified by the \emph{shared}
and \emph{global} channels to the top level family of such component.
Threads accessing objects in memory through a global reference are
the exception to this, but they have to be created on a specific exclusive
place in order to guarantee consistency.

Our strategy for building DSVP programs is based on the previous observations;
we can identify software components at the level of a family and its
children, that can be distributed to remote nodes with a \emph{create
}action using the corresponding \emph{place}. This component can then
internally create more threads locally on \emph{places} on that node,
or can decide at some point to create further sub components on other
nodes. However, the whole component has a single interface at its
top level family, and its dependencies are identified by the \emph{shared}
and \emph{global} channels to that family.

\paragraph{Distributed Memory}

As distribution is only done at the level of families, we can use
the information in the channels to the created family to determine
which data needs to be transferred. At the point of \emph{create,}
we synchronize or copy all objects that the family receives references
to, to the node it is created on. As all threads of the created family
run on the same place and therefore within the same consistent memory,
such replication is not required for internal communication of objects
between sibling threads. When the family completes, at the point of
\emph{sync}, they are synchronized or copied back again, taking into
account newly created references the family might send back through
its \emph{shared} channels. The second case where a family updates
global state on an exclusive place is not an issue; as each family
accessing this data is created on the same exclusive place, it shares
the same consistent memory, and no data communication is required
besides the earlier defined inputs and outputs.

This approach slightly restricts the original consistency model, as
it delivers consistency only for the memory areas that the child family
can effectively see. However, this approach is often too naive; for
example, it does not keep track of how data is used. Depending on
data being consumed or modified by the created family, we would like
to avoid copying back unmodified data for efficiency, so an implementation
has to detect or receive hints on which data has been modified. Furthermore,
on more complex large objects, e.g. a database, do we suffice with
a shallow copy or do we naively do an expensive deep copy of the object?
And what about objects with a non-static size? 

Some of these issues can be solved in a DSVP implementation or on top of that,
by using the notion of \emph{place} as we presented it for a distributed 
environment; instead of plain memory references, objects could be referenced 
by a combination of memory location and \emph{place}, as a place also 
identifies a memory range attached to a specific node. This way, a shallow
copy of complex objects is sufficient given that it internally uses this 
kind of \emph{fat} references, so that other referred objects can be fetched 
from the appropriate \emph{place} on demand. We decided not to make this 
mechanism part of our model for flexibility. DSVP already provides the 
necessary constructs so that this can be done on top of any implementation. 
Another observation is that an implementation would benefit from having more 
fine grained control over the inputs and outputs of a family, which
requires a programming language where we can either analyze or specify
in detail which data goes in, and which data is generated or modified
by a software component. In the next section we will discuss our prototype
implementation which uses a C based language, in which this analysis
is hard, and consequently we leave it to the programmer to explicitly specify
this. After all, the designer of a component has the best knowledge of what
its inputs and outputs are.

\section{Prototype Implementation over TCP/IP}\label{sec:Implementation}

We have built a prototype implementation of DSVP using the mechanisms
described in the previous section by extending the pthreads based
implementation of SVP~\cite{SVP-PTL09} with messages over TCP/IP
to signal the SVP actions between nodes. It supports heterogeneous
clusters of multi-core systems, connected with for example an Ethernet
network, where each system is a single node. This implementation is
driven by programs written in the C based $\mu$TC language \cite{uTC06},
in which threads are declared in a similar manner to C functions. Additional 
keywords are used to distinguish the shared and global channels in the 
arguments, but the input and output data is not explicitly indicated. This 
gives us the same problem as when attempting to analyze C functions; pointer
arguments may carry input data, output data, or both, and manipulation
of file-scope or global variables (side effects) is not indicated
at all. Yet, we must know exactly which data will need to be sent
to the remote place and back. Therefore, we require that the programmer,
or anything that generates $\mu$TC code, explicitly tells us what
the complete set of input and output data is in a \emph{data description
function}. Besides being a requirement for our implementation, this also
provides valuable documentation about the behaviour of a thread function.

\paragraph{Data Description Functions}

\begin{figure}[ht]
\begin{lstlisting}
DISTRIBUTABLE_THREAD(fibonacci)(int p1, 
             int p2, int* result, int N)
{
    INPUT(p1);
    INPUT(p2);
    for(int i = 2; i < N; i++)
    {
        OUTPUT(result[i]);
    }
}
\end{lstlisting}
\caption{\label{fig:Distributable-Fibonacci}Data description function for
Fibonacci}

\end{figure}
A data description function is a special function for each thread
function which describes the inputs and outputs using special
statements, allowing the corresponding thread function to be distributed
to other nodes by our DSVP implementation. This function receives the same 
arguments as the thread function, and is called by the implementation at
the creating and completing stage when the corresponding thread function 
is executed on a remote node. The data description function contains INPUT(v),
OUTPUT(v) and INOUT(v) statements, which trigger data transfers at the 
different stages. Data tagged with INPUT is copied to the remote node at the
stage when the thread function is started by a \emph{create}, and OUTPUT data 
is copied back to the creating node at the stage when the created family 
finishes and \emph{sync} completes. INOUT is a shorthand notation for the 
combination of the previous two.

Within these data description functions, loops and conditional expressions
can be used around the statements describing input and output. This provides
the flexibility needed in order to express the dynamic nature of family 
input/output data, for example dynamically sized arrays or the traversal of 
more complex data structures. In \prettyref{fig:Distributable-Fibonacci} 
we show how we can make the Fibonacci example code shown earlier in 
\prettyref{fig:Fibonacci-code} distributable by defining such a data 
description function. The startup values of the shared channels are only 
used as input to the Fibonacci function, and the array with the generated 
sequence is sent back as output. Please note that we needed to add the size
parameter to the thread function to support a non-fixed size for the
result array.

Using these data description functions we have a powerful way of expressing 
data dependencies and controlling which data goes into and comes out of a 
family of threads that is created on a remote node. Due to the restrictions on 
SVP programs, data only is communicated between two nodes at well known points, 
and no coherency protocols are required to keep data consistent. Because these 
data transfers are completely programmable in our prototype implementation, 
full control can be exercised over how data is distributed, for example for
splitting up arrays or array subsets across multiple nodes.

\paragraph{Types and Serialization}

For each thread function that needs to be distributable, the arguments
should consist of distributable data types, i.e. data types that the
implementation knows how to serialize and represent on the network.
Many standard C data-types are already provided as distributable,
but more complex objects such as structs or structs linked with pointers
must be defined using XDR \cite{XDR06}, which allows a syntax similar
to C. The XDR library provides us with (de)serialization, and guarantees
data interoperability between different architectures so that we can
support clusters of heterogeneous nodes. As long as a thread function
is defined to be distributable, it can be created both remotely and locally.
At run-time the implementation checks if a create is to a local or
a remote place, and only on a remote create will the (de)serialization
be performed; the function can still be created locally with
a negligible effect on performance compared to the non-distributed
implementation. In the distributed fibonacci example we've just shown,
we could have defined the results array as a new distributable data
type with a known fixed length, and then directly handle it with a
single OUTPUT() statement without the loop. Alternatively, it could
have been made dynamic by passing the length as an argument and then
using this in the loop bounds.

\paragraph{Message Implementation}

We have implemented a simple socket protocol over TCP/IP to send events
back and forth between nodes. The protocol consists of three
messages only;
\begin{itemize}
\item \textbf{create} -- is sent to the remote node and contains parameters
for the family to be created as well as the encoded input data.
\item \textbf{sync} (\emph{family finished}) -- is sent back from the remote
node on completion, it includes return values and the encoded output
data.
\item \textbf{kill} -- is sent to a remote node to interrupt the execution
of a family, it contains information to identify this family.
\end{itemize}
As we can see from this enumeration, the nature of the messages is
very simple, and induces minimal overhead. In general, the size of
the encoded in- or output data will be the dominant factor of the
message size. There is no message for the \emph{break} action, as
this it only applies to the family in which it is executed which will
always be on the same place, and when recursing to child families
it is the same as a recursing \emph{kill.} Our current implementation
does not contain any security, however this could be added easily
by introducing capabilities \cite{Capabilities66} on every message
and the use of \emph{places}, as well as by using encryption with
SSL sockets.

\section{Evaluation and Discussion}\label{sec:Evaluation}

\paragraph{Latency Measurement}

\begin{figure*}[ht]
\begin{center}
\includegraphics[angle=-90,origin=c,width=0.75\textwidth]{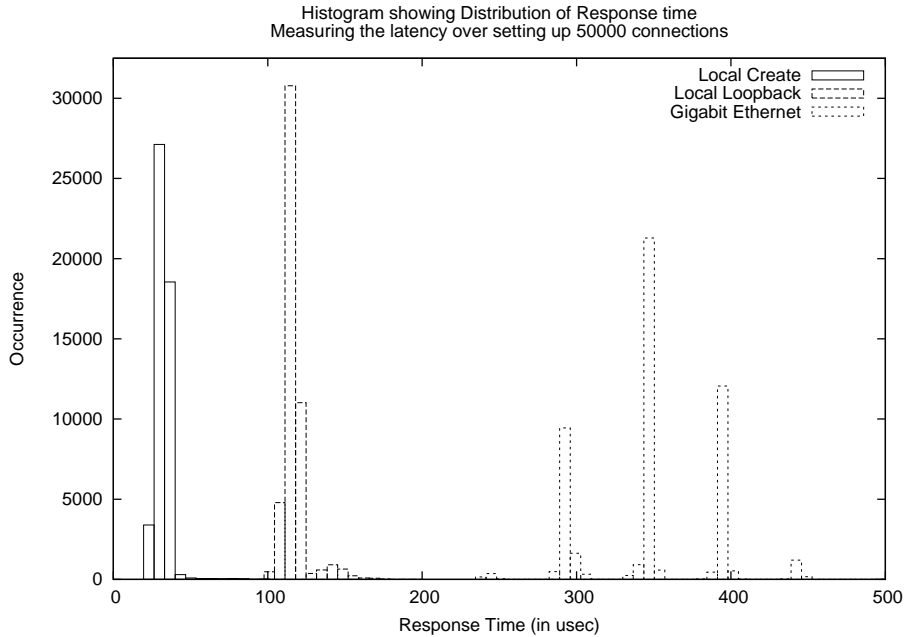}
\end{center}
\caption{\label{fig:Response-time}Response time of TCP/IP implementation}

\end{figure*}

We have measured the overhead imposed by our distributed implementation,
by measuring the latency over a paired \emph{create} and \emph{sync}
action on an empty thread function that executes remotely. For comparison,
we compare it with the latency of normal local creates of the same
function, as well as remote creates to a second runtime instance running
on the same machine over the internal loopback interface. These measurements
give us insight in the startup cost for remote executions, and allows us to 
make decisions about the level of granularity at which it is still feasible
to delegate to a different node when using this implementation. The
results of these measurements are shown as a histogram in \prettyref{fig:Response-time}
representing the distribution of latency over 50000 connections. These
experiments were all performed on Intel Dual-Core machines running
Linux kernel 2.6, which were connected with a direct non-switched
Gigabit Ethernet link. We see that creating a thread within another
process on the same machine using the local loopback on average takes 
$114\mu s$, and through Gigabit Ethernet it takes $345\mu s$ on average, 
but with $236\mu s$ as a minimum. The $50\mu s$ wide gaps between the peaks
that are observed in the Ethernet transmission are probably caused by an 
optimization in the
TCP/IP stack of the host systems that delays the delivery of ACK packets.
Not surprisingly, the overhead for creating threads over the network
is one order of magnitude slower than locally within the same runtime instance,
which are on average
created in around $30\mu s$. The difference between local create
and local loopback, is that in that case the whole protocol and serialization 
over a local TCP/IP socket is performed, as well as the scheduling between
two distinct processes.

\paragraph{Reducing Overhead}

In alternative implementations of the methods proposed in this report,
there is potential to greatly reduce the overhead compared to our prototype
implementation. For example, on a networked distributed system with
relatively homogeneous nodes, i.e. with the same internal data representation
(including endian-ness), time can be saved on (de)serialization and
encoding. On future many-core architectures or NUMA systems, the communication
between targets will likely use an efficient low overhead internal
messaging implementation instead of TCP/IP, and the same argument
against serialization holds. In fact, if we were to use such data
description functions, they would probably be used to synchronize
the data between the local and remote node by software coherency or
memory duplication. For the fully distributed heterogeneous platform
it could perhaps be beneficial to investigate a protocol based on
the much lighter UDP instead of the TCP/IP socket approach, as well
as a run-time that supports a finer grain of threads than pthreads.

\paragraph{Reference Transparency}

Many Distributed Shared Memory (DSM) implementations strive to have
a form of reference transparency where a reference to an object can
be accessed on any node. Usually this is done by using a shared address
space, and optionally using some special \emph{fat} references which
also encapsulate which node is the data's home location. As we have
argued earlier, such \emph{fat} references can be built on top
of DSVP as a combination between a \emph{place} and a normal reference,
and nothing prohibits a DSVP implementation from using a distributed
shared address space between nodes. However, with the prototype implementation
that we have described here, and using the restricted consistency
model of SVP, we achieve a similar programming model in a heterogeneous
environment. Also the implementation can support a dynamic number
of nodes, as the interaction between nodes is only limited to the
points where concurrency is created and synchronized. This is not
easy to achieve in a system that attempts to maintain a single global
address space.

\paragraph{Mapping and Resource Management}

Even though some notions of resources and their organization are visible
through the concept of \emph{places}, the extension of the SVP model
presented here is still resource agnostic. Using our prototype, we
can specify other nodes by hand so that they appear as \emph{places}
to the user program and can remotely execute functions. We have also
implemented a resource management system based on the SEP protocol
\cite{SEP08} which can do this dynamically. The details of that implementation
are beyond the scope of this report, but it supports the dynamic aggregation
of resources into a single DSVP system where nodes can join and leave
the network at run time, offering a set of software components as
services. A program can acquire a \emph{place} on another node and
then \emph{create} there one of the software components that the node
offers. As SVP only provides \emph{place} as a hook to identify resources
that a computation is bound to, and does not perform any mapping itself,
an SEP like service that acts as a \emph{place}-server handing out
\emph{places} could also take into account mapping and placing software
components efficiently.

\paragraph{Fault Tolerance}

Distributed systems have the disadvantage that communication is not
always reliable. The communication link, or perhaps even the whole
node might be unreliable or completely down. Besides distributed systems,
the many-core architectures of the future are not an exception to
these kinds of problems; with 1000s of cores on a chip it is unavoidable
that there will be faulty cores or interconnects present. Therefore,
it is essential that software on such platforms is fault tolerant
\cite{Wentzlaff2009,IntelTerascale07}.

In our implementation, we can use retries up to a certain level to
hide some of the communication problems, unless a target is not responding
within a reasonable amount of time. When waiting and retrying are
not enough, we want to inform the application, which then may give
up and display an error message, or could try to adapt itself to the
new situation. If the application is looking for generic resources
to execute a certain software component, it could try to get resources
to execute it on another target instead of the failed one. In terms
of SVP, this means sending a \emph{kill} to the old family that is
not responding, and creating a new family on a new place. As the input
and output data of a software component are defined, it can easily
be restarted on another target using the same input data again. Software
components that do maintain state at a place are typically services,
which are required to be implemented redundantly in such a system
using replication.

When the implementation cannot create a component on the desired target
or gets notified that the target failed, it will have the corresponding
\emph{sync} return an indication that the family did not complete
and the state of the output data is undefined. Similarly, if the component
fails to complete within an application-defined time, it can be killed
by a watchdog process, which is then reflected in the \emph{sync}
return value. In both cases, a new place should be selected, and the
component is (re)started there. This kind of flexibility in a system
can be very useful, not only to recover from communication errors
but also to adapt to, for example, dynamically changing load or availability
of resources.

\section{Related and Future Work}\label{sec:Related-Work}

Over the years, many ways of programming distributed environments
have been developed. There are distributed shared memory (DSM) implementations,
which for example use implicit or explicit sharing of objects \cite{Emerald88,Munin91,Orca93,Cid96},
regions \cite{CICO92,CRL95,CellSs06,Sequoia06}, or an entire address
space \cite{Ivy89}. The other end of the spectrum has been dominated
by explicit message passing techniques \cite{PVM94,MPI2-96}, and
in between we have remote calls (possibly to remote objects) \cite{RPC84,ActiveMessages92,CORBA06,JavaRMI96,Legion96,R-OSGi07},
which can also be based on web service interfaces \cite{SOAP1.1-2000}.
We will now discuss some of these approaches in more detail, and compare
them with DSVP.

Ivy \cite{Ivy89} was one of the first DSM systems that attempted
to act as a transparent single address space shared memory system
by sharing memory on the page level and using handlers on page miss
to transfer data. However, this did not turn out to work efficiently
enough, false sharing being one of the issues, and many later DSM
implementations are based on explicitly acquiring, reading or modifying
and releasing state. CRL \cite{CRL95} for example uses a region based
approach where special global pointers are used to map and unmap shared
regions of arbitrary size to code running on a node. After a region
is mapped, the code can enter either a reading or writing section,
where writing sections guarantee exclusive access. Munin \cite{Munin91}
also uses the acquire/release principle, but allows the consistency
protocol, which is based on release consistency \cite{ReleaseConsistency90},
to be configured for individual objects; i.e. invalidate or update
copies on write, enabling replication and fixed home locations. Cid
\cite{Cid96} also implements acquire/release with single writer multiple
readers, but also exposes the location of objects with the ability
to start a computation on an object on the node where it is located,
providing the flexibility of moving either the computation or the
data.

In Orca \cite{Orca93} the acquire/release happens transparently on
shared objects that get replicated. The objects are not globally visible
but are passed by reference between (concurrent) invocation of functions,
limiting their visibility to a relatively local scope similar as in
DSVP. However, when multiple functions operate on the same object
it is kept coherent by updating or invalidating copies on write. Emerald
\cite{Emerald88} provided similar mechanisms, however it did not
support replication and therefore did not have to deal with coherency.

CICO \cite{CICO92} is a cooperative model in which memory regions
in a shared address space can be checked out, in and prefetched, which
provides a hinting mechanism for a hardware based coherency implementation,
similarlar to how we see that the data description function annotations
could be used on a NUMA style system. This restricted way in which
we move data in and out of created families, has some similarities
and provides the same advantage as the DAG-consistency \cite{DAG-Consistency96}
provided in Cilk \cite{Cilk95}; in both there are well defined points when
data needs to be communicated, as there is no strict coherency which
requires propagation of updates as soon as data is modified. Another
approach that matches our work even more closely is CellSc \cite{CellSs06}
which uses compiler pragmas to annotate functions with their input
and output signature to efficiently write programs for the distributed
memory in the Cell \cite{CellBE06} architecture. Sequoia \cite{Sequoia06}
is a programming model in which a (distributed) system is viewed as
a hierarchy of memories, and, similar to SVP, programs in Sequioa
can be automatically adopted to the granularity of the target system.
Sequoia uses call-by-value-result semantics, where for each function
argument is specified if it describes an input, output or both. GMAC
\cite{GMAC10} is an implementation of an asynchronous distributed
shared memory which attempts to unify the programmability of CPU and
GPU memories. The Batch-update mode of GMAC matches closely with our
approach to consistency, however it also supports more elaborate coherency
protocols where the GPU can receive updated data from the CPU asynchronously.

In our definition of DSVP we unify the use of distributed memory and
dynamic concurrency management. Unifying the creation of local and
remote concurrency was investigated widely in the 90s, but was considered
a bad idea back then \cite{research.sun.com94}. This makes sense
as a remote execution on a cluster takes many orders of magnitude
more latency, and partial failure exposes different failure patterns.
However, we are on the brink of the many-core era and things have
changed. With many cores on one chip, starting an execution from one
core on another will be orders of magnitude faster than on a cluster.
And with thousands of cores on a chip, fault tolerance needs to be
supported to cater for failing cores and communication links \cite{IntelTerascale07,Wentzlaff2009}.
Checking for failure on any concurrent invocation would still be expensive,
but can be done at the software component level as discussed earlier.
R-OSGi \cite{R-OSGi07} is a system that takes this into account,
it distributes transparently at the software module level, and does
not introduce any new failure patterns. Similarly to our prototype
implementation, it does not impose any role assignments i.e. whether a
node acts as a client or server; the relation between modules is symmetric.
Chapel \cite{Chapel07} is a new programming language aimed to bridge
the gap between parallel and sequential programming. Similarly to
DSVP, it hierarchically expresses both task and data level concurrency,
which transparently can be executed locally or remotely in parallel, or
sequential, but it does not deal with partial failure. X10 \cite{X10-05}
is similar in that respect and is developed with the same goal as
Chapel. It bears more similarities to SVP with its \emph{futures}
and \emph{final} variables which resemble our shared and global channels.
It also uses places to express locations that have sequential consistency,
which provides a handle for expressing locality. Cid \cite{Cid96}
has this feature as well in a way, as the home node of a piece of
data can be extracted. This can then be used with its \emph{fork if
remote} construct, executing sequentially if the referenced object
is local, or otherwise remotely in parallel.

Other approaches such as Active Messages \cite{ActiveMessages92},
CORBA \cite{CORBA06}, Legion \cite{Legion96}, RPC \cite{RPC84},
Java RMI \cite{JavaRMI96} and SOAP \cite{SOAP1.1-2000} but also
message passing approaches such as MPI-2 \cite{MPI2-96} and PVM \cite{PVM94}
are based on coarse grained parallelism where finer grained
parallelism must be expressed in something else; for example in a
separate threading implementation. MPI-2 and PVM support the dynamic
creation of tasks, but again, only at task level parallelism. Most
of these approaches support partial failure, but at the cost of not
making remote communication transparent. None of them provide a distributed
memory abstraction, though CORBA, Java RMI and Legion do this in a
way by accessing remote objects. A lookup service is provided to locate
these objects, which can be added to DSVP by an SEP \cite{SEP08}
implementation.

Many of the discussed approaches rely on new languages or language
features, while others will work as pure library implementations.
DSVP does not exclude either of the two approaches; the prototype
implementation uses a C based language as input, but this is translated
to pure C++ with library calls \cite{SVP-PTL09} behind the scenes.
Of course, the argument for a language approach would be to be more
friendly or efficient for the programmer, but in our current toolchains
SVP is seen as an intermediate low level representation. There are
already tools to compile from SAC \cite{SAC06,SACSVP09}, a high level
array programming language, and an SVP based runtime for S-Net \cite{SNET2010,Graphwalker08},
a coordination language for streaming networks, has been developed.
As future work, we see that these tools can solve the problem of efficiently
describing the data dependencies as required for DSVP. These are well
known in the higher level representations of SAC and S-Net, and could
be automatically generated when compiling down to DSVP.

More future work lies in applying DSVP to emerging many-core architectures,
either in hardware or low-level software. We are currently working
on an implementation on the Intel SCC \cite{IntelSCC10}, an experimental
48-core processor created by Intel as a 'concept vehicle' platform
for many-core software research. This platform's NUMA style memory
organization and lack of cache coherence fit well with the distributed
style of memory for which DSVP was developed. We hope to exploit the
efficient on chip network for communication and delegation, as well
as its ability to change the memory mapping of each core.

\section{Conclusion}\label{sec:Conclusion}

In this report we have discussed how we can apply the SVP model of
concurrency to platforms with distributed memory organizations, which
is important in order to support decentralized memory organizations in 
future many-core architectures. We came to the conclusion that as long as 
we can identify software components and their data dependencies in 
SVP programs, we can trivially distribute them across multiple distributed 
memory domains. This approach fits the original memory consistency model of SVP
and still exposes the same restricted-consistency shared memory behavior.

We have discussed our prototype software implementation that we used 
to explore this domain. It can run SVP applications on TCP/IP networks of 
heterogeneous nodes, and uses data description functions to capture the 
dynamic nature of input and output data. We identified the minimal latencies 
imposed by this 
implementation to give an indication at which level of granularity it can be 
used efficiently. However, the main contribution are the techniques explored 
here that can be used as a basis for more fine grained SVP
implementations, applied to future or current many-core
architectures with distributed or non cache-coherent memory. As such
architectures and also distributed systems can suffer from partial 
failure, we have shown how the combination of the accurate description of 
input and output data and restricted points of communcation in SVP
can aid in the recovery of failure at the software component level.

\section{Acknowledgments}

The work presented in this report was based on work undertaken in the {\AE}THER
project, which was an IST-FET project funded by the European Community
under the FP6 ACA program. 

\bibliographystyle{abbrv}
\bibliography{references}

\begin{thebibliography}{10}

\bibitem{IntelTerascale07}
M.~Azimi, N.~Cherukuri, D.~N. Jayasimha, A.~Kumar, P.~Kundu, S.~Park,
  I.~Schoinas, and A.~S. Vaidya.
\newblock {Integration Challenges and Tradeoffs for Tera-scale Architectures}.
\newblock {\em Intel Technology Journal}, 11(3):173--184, 2007.

\bibitem{CellSs06}
P.~Bellens, J.~M. Perez, R.~M. Badia, and J.~Labarta.
\newblock Cellss: a programming model for the cell be architecture.
\newblock In {\em SC '06: Proceedings of the 2006 ACM/IEEE conference on
  Supercomputing}, page~86, New York, NY, USA, 2006. ACM.

\bibitem{RPC84}
A.~D. Birrell and B.~J. Nelson.
\newblock Implementing remote procedure calls.
\newblock {\em ACM Trans. Comput. Syst.}, 2:39--59, February 1984.

\bibitem{DAG-Consistency96}
R.~D. Blumofe, M.~Frigo, C.~F. Joerg, C.~E. Leiserson, and K.~H. Randall.
\newblock Dag-consistent distributed shared memory.
\newblock {\em Parallel Processing Symposium, International}, 0:132, 1996.

\bibitem{Cilk95}
R.~D. Blumofe, C.~F. Joerg, B.~C. Kuszmaul, C.~E. Leiserson, K.~H. Randall, and
  Y.~Zhou.
\newblock Cilk: an efficient multithreaded runtime system.
\newblock {\em SIGPLAN Not.}, 30(8):207--216, 1995.

\bibitem{Manycore07}
S.~Borkar.
\newblock Thousand core chips: a technology perspective.
\newblock In {\em DAC '07: Proceedings of the 44th annual Design Automation
  Conference}, pages 746--749, New York, NY, USA, 2007. ACM.

\bibitem{Microthread06}
K.~Bousias, N.~Hasasneh, and C.~Jesshope.
\newblock Instruction level parallelism through microthreading---a scalable
  approach to chip multiprocessors.
\newblock {\em Comput. J.}, 49:211--233, March 2006.

\bibitem{Graphwalker08}
K.~Bousias, C.~Jesshope, J.~Thiyagalingam, S.-B. Scholz, and A.~Shafarenko.
\newblock {Graph Walker: Implementing S-Net on the Self-adaptive Virtual
  Processor}.
\newblock In {\em Proceedings of the {\AE}ther-Morpheus Workshop From
  Reconfigurable to Self-Adaptive Computing (AMWAS'08)}, 2008.

\bibitem{SOAP1.1-2000}
D.~Box, D.~Ehnebuske, G.~Kakivaya, A.~Layman, N.~Mendelsohn, H.~{Frystyk
  Nielsen}, S.~Thatte, and D.~Winer.
\newblock Simple object access protocol (soap) 1.1.
\newblock World Wide Web Consortium, Note NOTE-SOAP-20000508, May 2000.

\bibitem{Munin91}
J.~B. Carter, J.~K. Bennett, and W.~Zwaenepoel.
\newblock Implementation and performance of munin.
\newblock In {\em SOSP '91: Proceedings of the thirteenth ACM symposium on
  Operating systems principles}, pages 152--164, New York, NY, USA, 1991. ACM.

\bibitem{Chapel07}
B.~Chamberlain, D.~Callahan, and H.~Zima.
\newblock Parallel programmability and the chapel language.
\newblock {\em Int. J. High Perform. Comput. Appl.}, 21(3):291--312, 2007.

\bibitem{X10-05}
P.~Charles, C.~Grothoff, V.~Saraswat, C.~Donawa, A.~Kielstra, K.~Ebcioglu,
  C.~von Praun, and V.~Sarkar.
\newblock X10: an object-oriented approach to non-uniform cluster computing.
\newblock In {\em OOPSLA '05: Proceedings of the 20th annual ACM SIGPLAN
  conference on Object-oriented programming, systems, languages, and
  applications}, pages 519--538, New York, NY, USA, 2005. ACM.

\bibitem{Capabilities66}
J.~B. Dennis and E.~C. Van~Horn.
\newblock Programming semantics for multiprogrammed computations.
\newblock {\em Commun. ACM}, 9(3):143--155, 1966.

\bibitem{XDR06}
M.~Eisler.
\newblock {XDR: External Data Representation Standard}.
\newblock RFC 4506 (Standard), May 2006.

\bibitem{Sequoia06}
K.~Fatahalian, D.~R. Horn, T.~J. Knight, L.~Leem, M.~Houston, J.~Y. Park,
  M.~Erez, M.~Ren, A.~Aiken, W.~J. Dally, and P.~Hanrahan.
\newblock Sequoia: programming the memory hierarchy.
\newblock In {\em SC '06: Proceedings of the 2006 ACM/IEEE conference on
  Supercomputing}, page~83, New York, NY, USA, 2006. ACM.

\bibitem{PVM94}
A.~Geist, A.~Beguelin, J.~Dongarra, W.~Jiang, R.~Manchek, and V.~Sunderam.
\newblock {\em PVM: Parallel virtual machine: a users' guide and tutorial for
  networked parallel computing}.
\newblock MIT Press, Cambridge, MA, USA, 1994.

\bibitem{MPI2-96}
A.~Geist, W.~Gropp, S.~Huss-Lederman, A.~Lumsdaine, E.~Lusk, W.~Saphir,
  T.~Skjellum, and M.~Snir.
\newblock Mpi-2: Extending the message-passing interface.
\newblock In L.~Bougé, P.~Fraigniaud, A.~Mignotte, and Y.~Robert, editors, {\em
  Euro-Par'96 Parallel Processing}, volume 1123 of {\em Lecture Notes in
  Computer Science}, pages 128--135. Springer Berlin / Heidelberg, 1996.

\bibitem{GMAC10}
I.~Gelado, J.~E. Stone, J.~Cabezas, S.~Patel, N.~Navarro, and W.-m.~W. Hwu.
\newblock An asymmetric distributed shared memory model for heterogeneous
  parallel systems.
\newblock In {\em ASPLOS '10: Proceedings of the fifteenth edition of ASPLOS on
  Architectural support for programming languages and operating systems}, pages
  347--358, New York, NY, USA, 2010. ACM.

\bibitem{ReleaseConsistency90}
K.~Gharachorloo, D.~Lenoski, J.~Laudon, P.~Gibbons, A.~Gupta, and J.~Hennessy.
\newblock Memory consistency and event ordering in scalable shared-memory
  multiprocessors.
\newblock In {\em ISCA '90: Proceedings of the 17th annual international
  symposium on Computer Architecture}, pages 15--26, New York, NY, USA, 1990.
  ACM.

\bibitem{SACSVP09}
C.~Grelck, S.~Herhut, C.~Jesshope, C.~Joslin, M.~Lankamp, S.-B. Scholz, and
  A.~Shafarenko.
\newblock {Compiling the Functional Data-Parallel Language SAC for Microgrids
  of Self-Adaptive Virtual Processors}.
\newblock In {\em 14th Workshop on Compilers for Parallel Computing (CPC'09),
  IBM Research Center, Z\"urich, Switzerland}, 2009.

\bibitem{SAC06}
C.~Grelck and S.-B. Scholz.
\newblock {SAC}: A functional array language for efficient multithreaded
  execution.
\newblock {\em International Journal of Parallel Programming}, 34(4):383--427,
  2006.

\bibitem{SNET2010}
C.~Grelck, {Shafarenko, A. (eds):}, F.~Penczek, C.~Grelck, H.~Cai, J.~Julku,
  P.~H\"olzenspies, {Scholz, S.B.}, and A.~Shafarenko.
\newblock {S-Net Language Report 2.0}.
\newblock Technical Report 499, University of Hertfordshire, School of Computer
  Science, Hatfield, England, United Kingdom, 2010.

\bibitem{CORBA06}
O.~M. Group.
\newblock Corba component model 4.0 specification.
\newblock Specification Version 4.0, Object Management Group, April 2006.

\bibitem{CellBE06}
M.~Gschwind, H.~Hofstee, B.~Flachs, M.~Hopkin, Y.~Watanabe, and T.~Yamazaki.
\newblock Synergistic processing in cell's multicore architecture.
\newblock {\em Micro, IEEE}, 26(2):10 --24, march-april 2006.

\bibitem{Heiser2009}
G.~Heiser.
\newblock Many-core chips -- a case for virtual shared memory.
\newblock In {\em Proceedings of the 2nd Workshop on Managed Many-Core
  Systems}, Washington, DC, USA, March 2009. ACM.

\bibitem{SVPIO10}
M.~A. Hicks, M.~W. van Tol, and C.~R. Jesshope.
\newblock Towards scalable i/o on a many-core architecture.
\newblock In {\em Embedded Computer Systems (SAMOS), 2010 International
  Conference on}, pages 341--348, 2010.

\bibitem{CICO92}
M.~D. Hill, J.~R. Larus, S.~K. Reinhardt, and D.~A. Wood.
\newblock Cooperative shared memory: software and hardware for scalable
  multiprocessor.
\newblock In {\em ASPLOS-V: Proceedings of the fifth international conference
  on Architectural support for programming languages and operating systems},
  pages 262--273, New York, NY, USA, 1992. ACM.

\bibitem{IntelSCC10}
J.~Howard, S.~Dighe, Y.~Hoskote, S.~Vangal, D.~Finan, G.~Ruhl, D.~Jenkins,
  H.~Wilson, N.~Borkar, G.~Schrom, F.~Pailet, S.~Jain, T.~Jacob, S.~Yada,
  S.~Marella, P.~Salihundam, V.~Erraguntla, M.~Konow, M.~Riepen, G.~Droege,
  J.~Lindemann, M.~Gries, T.~Apel, K.~Henriss, T.~Lund-Larsen, S.~Steibl,
  S.~Borkar, V.~De, R.~van~der Wijngaart, and T.~Mattson.
\newblock A 48-core ia-32 message-passing processor with dvfs in 45nm cmos.
\newblock In {\em Solid-State Circuits Conference - Digest of Technical Papers,
  2010. ISSCC 2010. IEEE International}, pages 19--21, February 2010.

\bibitem{SVP08}
C.~Jesshope.
\newblock {A model for the design and programming of multi-cores}.
\newblock {\em Advances in Parallel Computing}, High Performance Computing and
  Grids in Action(16):37--55, 2008.

\bibitem{Microgrid09}
C.~Jesshope, M.~Lankamp, and L.~Zhang.
\newblock {Evaluating CMPs and their memory architecture}.
\newblock In M.~Berekovic, C.~Muller-Schoer, C.~Hochberger, and S.~Wong,
  editors, {\em Proc. Architecture of Computing Systems}, pages 246--257, 2009.

\bibitem{SEP08}
C.~Jesshope, J.-M. Philippe, and M.~van Tol.
\newblock An architecture and protocol for the management of resources in
  ubiquitous and heterogeneous systems based on the svp model of concurrency.
\newblock In {\em SAMOS '08: Proceedings of the 8th international workshop on
  Embedded Computer Systems}, pages 218--228, Berlin, Heidelberg, 2008.
  Springer-Verlag.

\bibitem{uTC06}
C.~R. Jesshope.
\newblock $\mu$tc - an intermediate language for programming chip
  multiprocessors.
\newblock In C.~R. Jesshope and C.~Egan, editors, {\em Asia-Pacific Computer
  Systems Architecture Conference}, volume 4186 of {\em Lecture Notes in
  Computer Science}, pages 147--160. Springer, 2006.

\bibitem{CRL95}
K.~L. Johnson, M.~F. Kaashoek, and D.~A. Wallach.
\newblock Crl: high-performance all-software distributed shared memory.
\newblock In {\em SOSP '95: Proceedings of the fifteenth ACM symposium on
  Operating systems principles}, pages 213--226, New York, NY, USA, 1995. ACM.

\bibitem{Emerald88}
E.~Jul, H.~Levy, N.~Hutchinson, and A.~Black.
\newblock Fine-grained mobility in the emerald system.
\newblock {\em ACM Trans. Comput. Syst.}, 6:109--133, February 1988.

\bibitem{Kurian2010}
G.~Kurian, N.~Beckmann, J.~Miller, J.~Psota, and A.~Agarwal.
\newblock Efficient cache coherence on manycore optical networks.
\newblock Technical Report MIT-CSAIL-TR-2010-009, Computer Science and
  Artificial Intelligence Lab, MIT, February 2010.

\bibitem{Legion96}
M.~Lewis and A.~Grimshaw.
\newblock The core legion object model.
\newblock In {\em HPDC '96: Proceedings of the 5th IEEE International Symposium
  on High Performance Distributed Computing}, page 551, Washington, DC, USA,
  1996. IEEE Computer Society.

\bibitem{Ivy89}
K.~Li and P.~Hudak.
\newblock Memory coherence in shared virtual memory systems.
\newblock {\em ACM Trans. Comput. Syst.}, 7:321--359, November 1989.

\bibitem{Cid96}
R.~Nikhil.
\newblock {\em Parallel symbolic computing in Cid}, volume 1068 of {\em Lecture
  Notes in Computer Science}, pages 215--242.
\newblock Springer-Verlag Berlin / Heidelberg, 1996.

\bibitem{R-OSGi07}
J.~S. Rellermeyer, G.~Alonso, and T.~Roscoe.
\newblock R-osgi: distributed applications through software modularization.
\newblock In {\em Middleware '07: Proceedings of the ACM/IFIP/USENIX 2007
  International Conference on Middleware}, pages 1--20, New York, NY, USA,
  2007. Springer-Verlag New York, Inc.

\bibitem{Orca93}
A.~Tanenbaum, H.~Bal, and M.~Kaashoek.
\newblock Programming a distributed system using shared objects.
\newblock In {\em High Performance Distributed Computing, 1993., Proceedings
  the 2nd International Symposium on}, pages 5 --12, 20-23 1993.

\bibitem{SVP-PTL09}
M.~van Tol, C.~Jesshope, M.~Lankamp, and S.~Polstra.
\newblock An implementation of the sane virtual processor using posix threads.
\newblock {\em Journal of Systems Architecture}, 55(3):162--169, 2009.
\newblock Challenges in self-adaptive computing (Selected papers from the
  Aether-Morpheus 2007 workshop).

\bibitem{ActiveMessages92}
T.~von Eicken, D.~E. Culler, S.~C. Goldstein, and K.~E. Schauser.
\newblock Active messages: a mechanism for integrated communication and
  computation.
\newblock In {\em ISCA '92: Proceedings of the 19th annual international
  symposium on Computer architecture}, pages 256--266, New York, NY, USA, 1992.
  ACM.

\bibitem{FormalSVP07}
T.~D. Vu and C.~R. Jesshope.
\newblock Formalizing sane virtual processor in thread algebra.
\newblock In M.~Butler, M.~G. Hinchey, and M.~M. Larrondo-Petrie, editors, {\em
  ICFEM}, volume 4789 of {\em Lecture Notes in Computer Science}, pages
  345--365. Springer, 2007.

\bibitem{research.sun.com94}
J.~Waldo, G.~Wyant, A.~Wollrath, and S.~C. Kendall.
\newblock A note on distributed computing.
\newblock Technical Report TR-94-29, Sun Microsystems Research, November 1994.

\bibitem{Wentzlaff2009}
D.~Wentzlaff, C.~Gruenwald, N.~Beckmann, K.~Modzelewski, A.~Belay, L.~Youseff,
  J.~Miller, and A.~Agarwal.
\newblock A unified operating system for clouds and manycore: fos.
\newblock Technical Report MIT-CSAIL-TR-2009-059, Computer Science and
  Artificial Intelligence Lab, MIT, November 2009.

\bibitem{JavaRMI96}
A.~Wollrath, R.~Riggs, and J.~Waldo.
\newblock A distributed object model for the java system.
\newblock In {\em COOTS}. USENIX, 1996.

\end{thebibliography}

\end{document}